\pdfoutput=1
\documentclass[twocolumn,english,aps,superscriptaddress,prb]{revtex4-1}

\usepackage{babel}
\usepackage{amsmath}
\usepackage{amssymb}
\usepackage{graphicx}

\begin{document}

\title{Rigorous decoupling between edge states in frustrated spin chains and ladders }

\author{Natalia Chepiga}
\affiliation{Department of Physics and Astronomy, University of California, Irvine, CA 92697, USA}
\author{Fr\'ed\'eric Mila}
\affiliation{Institute of Physics, Ecole Polytechnique F\'ed\'erale de Lausanne (EPFL), CH-1015 Lausanne, Switzerland}

\date{\today}
\begin{abstract} 
We investigate the occurrence of exact zero modes in one-dimensional quantum magnets of finite length that possess edge states. Building on conclusions first reached in the context of the spin-1/2 XY chain in a field, then for the spin-1 $J_1-J_2$ Heisenberg model, we show that the development of incommensurate correlations in the bulk invariably leads to oscillations in the sign of the coupling between edge states, hence to exact zero energy modes at the crossing points where the coupling between the edge states rigorously vanishes. This is true regardless of the origin of the frustration (e.g. next-nearest neighbor coupling or biquadratic coupling for the spin-1 chain), of the value of the bulk spin (we report on spin-1/2, spin-1 and spin-2 examples), and of the value of the edge-state emergent spin (spin-1/2 or spin-1). 
\end{abstract}
\pacs{
75.10.Jm,75.10.Pq,75.40.Mg
}

\maketitle


\section{Introduction}

Topological states of matter are currently attracting a lot of attention.\cite{RevModPhys.82.3045,RevModPhys.83.1057} In many contexts topologically non-trivial states are associated with the appearance of edge states. Majorana fermions appear at the edges of the Kitaev chain in the topologically non-trivial phase and can be detected by the presence of two quasi-degenerate low-lying states.\cite{kitaev,mourik,nadj-perge,dassarma}. Recently it has been shown that these two states might cross as a function of an external parameter such as the chemical potential\cite{dassarma}. Such level crossings have been recently detected in chains of Co adatoms\cite{toskovic}. These chains are described by an effective spin-1/2 XY model in a field, a model that can be mapped on the Kitaev chain by a Jordan-Wigner transformation, and the interpretation of the level crossings observed as a function of the field in terms of localized
Majorana fermions has been worked out in details\cite{mila,vionnet}. At each level crossing, there is an exact zero mode, i.e. an excitation whose energy vanishes exactly. In the fermionic model, the exact zero modes appear when the Majorana edge states are rigorously decoupled.

Another well known example of topological quantum states is the spin-1 Heisenberg chain, which has long been known to have a finite bulk gap\cite{haldane} and spin-1/2 edge states\cite{kennedy,hagiwara}. In the Heisenberg spin-1 chain, these spin-1/2 edge states form two quasi-degenerate low-lying states, a singlet and a triplet (the Kennedy triplet \cite{kennedy,hagiwara}), and the energy gap between them decays exponentially with the length of the chain. Recently it has been shown that the effective coupling between the spin-1/2 edge states can be continuously tuned by frustration\cite{chepiga_zero_modes}. For the $J_1-J_2$ model with nearest and next-nearest-neighbor antiferromagnetic interactions the singlet and the triplet low-lying states cross several times in the parameter range $0.28\lesssim J_2/J_1 \lesssim 0.75$, between the disorder point and the first-order transition to the next-nearest neighbour Haldane phase.

In both cases, the level crossings are intimately connected with the development of incommensurate fluctuations. In the XY model in a transverse field, the spin-spin correlations are incommensurate up to the saturation field. More generally for the anisotropic version of the model with different couplings in the $x$ and $y$ directions, the spin-spin correlations are incommensurate in a field range that shrinks to zero only at the Ising point, and the level crossings all occur in the field range where the spin-spin correlations are incommensurate\cite{vionnet}. In the case of the spin-1 $J_1-J_2$ chain, an even more direct connection has been established\cite{chepiga_zero_modes}. Indeed, it has been demonstrated that the sign of the effective coupling between the spin-1/2 edge spins follows to a good accuracy the sign of spin-spin correlations between the first and last spins-1 of the chain, and that the sign changes are themselves a direct consequence of the incommensurate fluctuations: the sign changes roughly as $\cos(qL)$, where $q$ is the wave-vector of the incommensurate fluctuations and $L$ the length of the chain, so that, for given parameter, hence for a given $q$, the sign of the coupling oscillates as a function of the chain length.

In the present paper we go further in the study of the appearance of exact zero modes in frustrated one-dimensional spin systems. 
We show that level crossings appear between low-lying in-gap states in a large variety of models, including models with spin-1 edge states, to reach the conclusion that the appearance of exact zero modes is a generic feature of systems with incommensurate correlations and localized edge states. 

The rest of the paper is organized as follows. In Section \ref{sec:bb} we study the appearance of exact zero modes in the spin-1 chain with bilinear-biquadratic interaction. The model is known to be in the incommensurate regime of the Haldane phase between the Affleck-Kennedy-Lieb-Tasaki (AKLT) point and the critical WZW SU$(3)$ point. Section \ref{sec:ladder} is devoted to the frustrated spin-1/2 ladder with diagonal edges, and section \ref{sec:spin2} to the antiferromagnetic $J_1-J_2$ spin-2 chain with localized spin-1 edge states. In Section \ref{sec:fms1}, we show that localized spin-1 edge states are also present in the $J_1-J_2$ spin-1 chain if the nearest-neighbor coupling is ferromagnetic ($J_1<0$). The results are briefly summarized in Section \ref{sec:conclusion}.


\section{Methods}

Two types of numerical simulations have been used. For the spin-1/2 ladder, we have performed exact diagonalizations using Lanczos algorithm. For the frustrated spin chains with $S\geq 1$, all the results have been obtained with the Density matrix Renormalization Group (DMRG) algorithm\cite{dmrg1,dmrg2,dmrg3,dmrg4}. In order to compute the energy of several in-gap states we have targetted multiple states in two-site DMRG\cite{chepiga_dmrg}. The method turns out to be extremely precise, and in the vicinity of the disorder point it provides the energy splitting between in-gap states with machine precision. Far from the disorder points, where the correlation length is larger and the convergence becomes slower, we have kept up to 1200 states (1500 for spin-2). In all cases, this has allowed us to extract the energy with an error below $10^{-10}$.

\section{Bilinear-biquadratic spin-1 chain}
\label{sec:bb}

\begin{figure}[t!]
\includegraphics[width=0.49\textwidth]{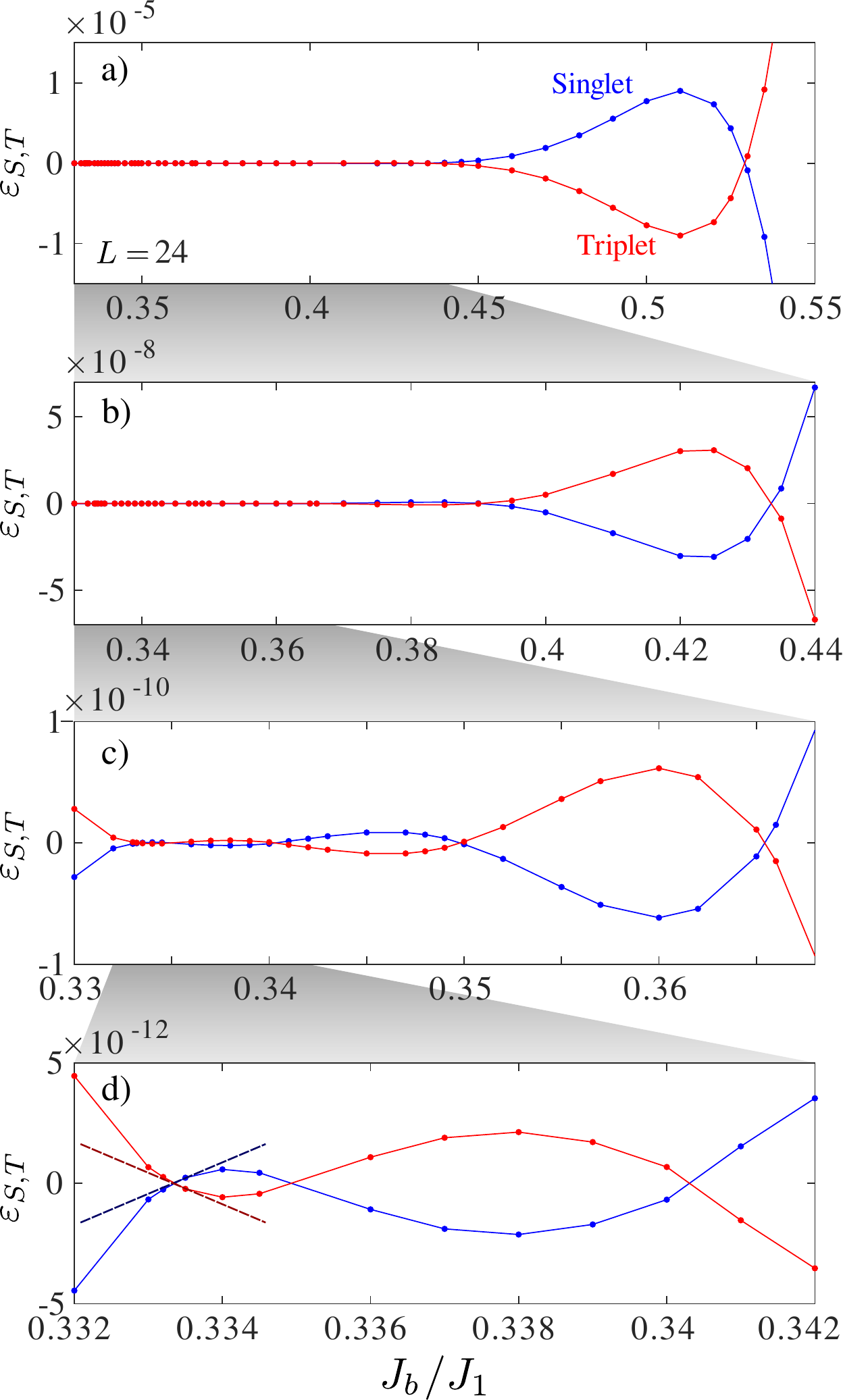}
\caption{(Color online) Multiple crossings between the singlet and triplet low-lying energy levels of the bilinear-biquadratic spin-1 chain of Eq.(\ref{eq:bilinbiquadr}) for L=24 as a function of the biquadratic coupling constant $J_b$. (b), (c) and (d) are enlarged parts of (a). }
\label{fig:energy_bilinbiquadr}
\end{figure}

An important milestone in the confirmation of Haldane's prediction of a finite bulk gap in the spin-1 chain was the construction of an exact ground state known as the Affleck-Kennedy-Lieb-Tasaki (AKLT) state. This state is represented by single valence bonds connecting each nearest-neighbor pair of spins. The parent Hamiltonian for which the AKLT state is an exact ground state  is given by the spin-1 bilinear-biquadratic model:

\begin{equation}
H=\sum_{i=1}^{L-1}J_1{\bf S}_{i}\cdot{\bf S}_{i+1}+J_b\left({\bf S}_{i}\cdot{\bf S}_{i+1}\right)^2,
\label{eq:bilinbiquadr}
\end{equation}
with $J_b/J_1=1/3$, a case also known as the AKLT point. At this point, the ground state wave-functions contains two completely decoupled edge spins, and accordingly the singlet and triplet low-lying states are exactly degenerate whatever the length of the chain. The AKLT point turns out to be also a disorder point \cite{schollwoeck_bilbiq}, i.e. a point beyond which the correlations are incommensurate. At $J_b/J_1=1$, the system undergoes a continuous Wess-Zimino-Witten (WZW) SU$(3)$ phase transition into a critical antiferroquadrupolar phase \cite{fath_solyom}. 
So the system is in the Haldane phase with localized edge excitations and incommensurate correlations for $1/3\leq J_b/J_1\leq 1$. Within this parameter range, we have detected multiple crossings between the singlet and the triplet low-lying in-gap states as shown in Fig.\ref{fig:energy_bilinbiquadr}. This feature of the bilinear-biquadratic spin-1 chain has been reported previously on small clusters with $L=10$ spins\cite{nomura}. For convenience, the energies have been rescaled according to  $\varepsilon_{S,T}=E_{S,T}-(E_S+E_T)/2$.

For any system size, the first crossing takes place exactly at the AKLT point. This absence of finite-size effect is due to the fact that the AKLT point is an exactly solvable point at which the emergent spins-1/2 are completely decoupled for any system size. By contrast, in the $J_1-J_2$ model studied previously\cite{chepiga_zero_modes}, the position of the first crossing point slightly deviates in small systems from the disorder point (defined in the thermodynamic limit). 
Interestingly, the coincidence between the point where the ground-state is an exact product of singlets and the disorder point where correlations become incommensurate also occurs in other  models. For instance, in the dimerized phase of the $J_1-J_2$ spin-1/2 chain, disorder develops exactly at the Majumdar-Ghosh point.  For the generalization of the Majumdar-Ghosh point to higher spins, which requires an additional three-site interaction $J_3$ \cite{michaud1, michaud2}, the fully dimerized state is an exact ground-state along a line $J_3/(J_1-2J_2)=1/\left[4S(S+1)-2\right]$ in the $J_1-J_2-J_3$ parameter space\cite{wang}. For the spin-1 case, it has been shown that the disorder line again coincides exactly with the fully dimerized line\cite{chepigaPRB}, and we expect that this remains true for higher spins.

As a confirmation of the numerical results of Fig.\ref{fig:energy_bilinbiquadr}, let us calculate the slopes of the singlet and triplet ground states of the AKLT point away from that point using their explicit form. The ground states at the AKLT point can be written in a simple and exact way using matrix product state (MPS) tensors\cite{dmrg4} with an auxiliary bond dimension $D=2$. Let us briefly remind the construction of the MPS for the AKLT state. One starts with $2N$ spins-1/2, that are completely symmetrized on every second bond to form a triplet that is identified with a spin-1:

\begin{equation}
\begin{array}{lcl}
  |t^+\rangle=|\uparrow\uparrow\rangle \\ \\
  |t^0\rangle=\frac{|\uparrow\downarrow\rangle+|\downarrow\uparrow\rangle}{\sqrt{2}} \\ \\
  |t^-\rangle=|\downarrow\downarrow\rangle
  \end{array}
\end{equation} 
Spin-1/2 states $|\uparrow\rangle$ and $|\downarrow\rangle$ can be considered as a basis for auxiliary indices $a$ and $b$. The on-site tensor $T_{a,b}^\sigma$ with physical index $\sigma$ that has dimension $d=3$ and corresponds to spin-1 is given by:
 \begin{equation}
 \begin{array}{lcl}
  T^{\sigma=1}=  \begin{bmatrix}
    1 & 0\\
    0 & 0
  \end{bmatrix}\\ \\
  T^{\sigma=0}=  \begin{bmatrix}
    0 & \frac{1}{\sqrt{2}}\\
    \frac{1}{\sqrt{2}} & 0
  \end{bmatrix}\\ \\
  T^{\sigma=-1}=  \begin{bmatrix}
    0 & 0\\
    0 & 1
  \end{bmatrix}
  \end{array}
\end{equation} 

\begin{figure}[t!]
\includegraphics[width=0.49\textwidth]{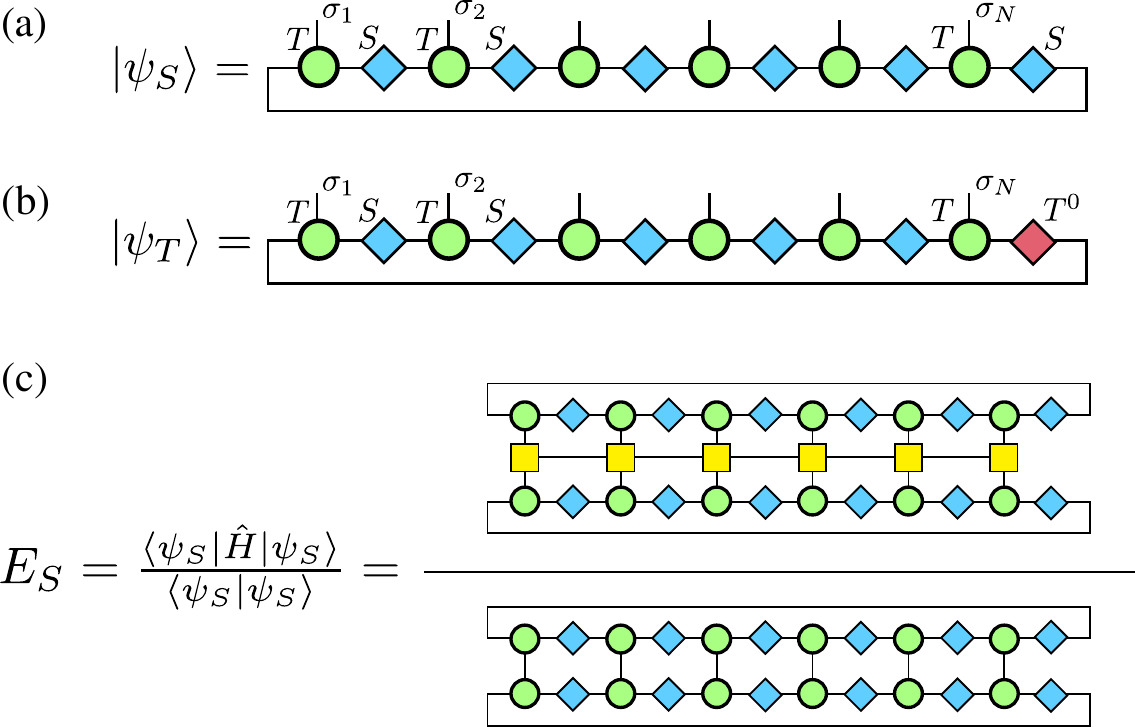}
\caption{(Color online) Sketch of the tensor network that represents an AKLT state with (a) singlet and (b) triplet coupling between the edge spins. Green circles represent an on-site tensor with physical index $\sigma_i$ of dimension $d=3$ that corresponds to a spin-1. Blue diamonds represent a projector onto a singlet state; red diamonds represent a projector onto one of the triplet states (here $T^\sigma=0$). Connecting lines correspond to the tensors contraction over the corresponding bonds.  (c) Graphical representation of the tensor network contraction to compute the energy of the singlet in-gap state. Yellow boxes represent the bilinear-biquadratic Hamiltonian of Eq.(\ref{eq:bilinbiquadr}) written in terms of on-site matrix product operators. The generalization to the energy of the triplet state is straightforward.  }
\label{fig:AKLT_MPS}
\end{figure}

On every other bond, two spins-1/2 form a singlet:
\begin{equation}
    |s\rangle=\frac{|\uparrow\downarrow\rangle-|\downarrow\uparrow\rangle}{\sqrt{2}}
\end{equation}
Therefore on-site tensors $T_{a,b}^\sigma$ are contracted with each other through a {\it bond tensor} without a physical index:
\begin{equation}
  S=  \begin{bmatrix}
    0 & \frac{1}{\sqrt{2}}\\
    -\frac{1}{\sqrt{2}} & 0
  \end{bmatrix}
\end{equation}
The tensor network representation of the AKLT state with open boundary conditions can be written as:
\begin{equation}
  T^{\sigma_1}_{a1,b1}S_{b1,a2}T^{\sigma_2}_{a2,b2}S_{b2,a3}... S_{b_{N-1},a_N}T^{\sigma_N}_{a_N,b_N},
\end{equation}
where all repeated indices are summed over. The sum does not run over indices  $a_1$ and $b_N$. This results in a $2\times 2$ matrix written in the basis of the edge spins-1/2. The singlet in-gap state can be obtained by projecting the edge spins onto an anti-symmetric state with the help of $S$ matrix. The corresponding wave-function is given by:
\begin{equation}
  |\psi_S\rangle=T^{\sigma_1}_{a1,b1}S_{b1,a2}T^{\sigma_2}_{a2,b2}S_{b2,a3}... T^{\sigma_N}_{a_N,b_N}S_{b_N,a_1}
  \label{eq:psi_s}
\end{equation}
The graphical representation of this tensor network is sketched in Fig.\ref{fig:AKLT_MPS}(a), where connecting lines represent the contraction of the tensors. 

The Kennedy triplet state can be obtained by projecting the two edge spins onto one of the three triplet states, e.g. by inserting the $T^{\sigma=0}_{a,b}$ matrix between the first and last sites, as shown in Fig.\ref{fig:AKLT_MPS}(b). The corresponding wave-function is given by the following matrix product:
\begin{equation}
  |\psi_T\rangle=T^{\sigma_1}_{a1,b1}S_{b1,a2}T^{\sigma_2}_{a2,b2}S_{b2,a3}... T^{\sigma_N}_{a_N,b_N}T^0_{b_N,a_1}
  \label{eq:psi_t}
\end{equation}

Note that the wave-functions $|\psi_S\rangle$ and $|\psi_T\rangle$ are not normalized. This has to be taken into account when computing the energy of each state as shown in Fig.\ref{fig:AKLT_MPS}(c). 

At the AKLT point, since the two edge spins are completely decoupled, singlet and triplet states are exactly degenerate.  
We have calculated the slopes of these energy levels  around the AKLT point by contracting the exact MPS given by Eq.(\ref{eq:psi_s}) and (\ref{eq:psi_t}) with the Hamiltonian written in the vicinity of the AKLT point $J_b=1/3\pm\varepsilon$, where $\varepsilon\ll 1$ is encoded as a symbolic variable. The slopes of the singlet and triplet gap match our DMRG data around the AKLT point (see Fig.\ref{fig:energy_bilinbiquadr}(d)). Note that since the correlation length is extremely small around the AKLT point and since our numerical method allows one to detect a gap only if it exceeds the machine precision $10^{-16}$, it is only possible to detect {\bf all} the level crossings on relatively small clusters with $L  \lesssim 26$. 

To summarize, the apparition of exact gapless points in the frustrated Haldane chain is independent from the type of frustration as long as it leads to incommensurate correlations within the Haldane phase. If the disorder point is exact (i.e size independent), the first level crossing always occurs at this point. 


\section{Spin-1/2 ladder with diagonal edges}
\label{sec:ladder}

\begin{figure}[h!]
\includegraphics[width=0.45\textwidth]{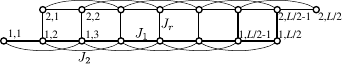}
\caption{(Color online) Spin-1/2 two-leg ladder with nearest- and next-nearest-neighbor intra-chain interaction and diagonal edges. The site indices correspond to the convention of Eq.\ref{eq:ladder}.}
\label{fig:modlad}
\end{figure}

 In recent years, the investigation of the topological properties spin-1/2  ladders has  been  a very active field of research.\cite{PhysRevB.79.115107,PhysRevB.91.214410} It has long been known that some topologically non-trivial states can be revealed by the presence of  localized edge states that appear in two-leg ladders with diagonal edges but are absent in the case of vertical edges\cite{PhysRevB.79.205107,PhysRevB.88.184418}. 
 In this section we consider the antiferromagnetic spin-1/2 ladder with an additional next-nearest-neighbor interaction along the legs (see Fig.\ref{fig:modlad}) that induces incommensurate correlations.
 
 The system is described by the following Hamiltonian:

\begin{multline}
H=\sum_{\alpha=1,2}\left[\sum_{i=1}^{L/2-1}J_1{\bf S}_{\alpha,i}\cdot{\bf S}_{\alpha,i+1}+\sum_{i=1}^{L/2-2}J_2{\bf S}_{\alpha,i}\cdot{\bf S}_{\alpha,i+2}\right]\\
+\sum_{i=2}^{L/2}J_r{\bf S}_{1,i}\cdot{\bf S}_{2,i-1},
\label{eq:ladder}
\end{multline}
where $J_1$ and $J_2$ are nearest and next-nearest-neighbor intra-chain couplings, and $J_r$ is the inter-chain coupling (see Fig.\ref{fig:modlad}); $L$ is a total number of spins. The following parametrization is used for convenience: $J_1=\cos\theta$, $J_2=j_2\cos\theta$ and $J_r=\sin\theta$. When $\theta=0$ the system corresponds to two decoupled $J_1-J_2$ spin-1/2 chains. When $\theta=\pi/2$, the intra-chain coupling is absent and the system corresponds to the product of rung singlets.

The correlations are incommensurate beyond the disorder line that starts at the Majumdar-Ghosh point $j_2=1/2$ and $\theta=0$ and goes up to the point $\theta=\pi/2$, $J_2/J_1=0$. The location of the disorder line between these two points has been determined by looking at the emergent incommensurability in the spin-spin correlations $C(x)=\langle S^z_{L/2}S^z_{L/2+x}\rangle$ and at the kink of the correlation length\cite{schollwoeck_bilbiq,garel}. For any finite $j_2$ the line $\theta=\pi/2$ corresponds to the exact rung dimer state and thus coincides with the second disorder line.

\begin{figure}[h!]
\includegraphics[width=0.47\textwidth]{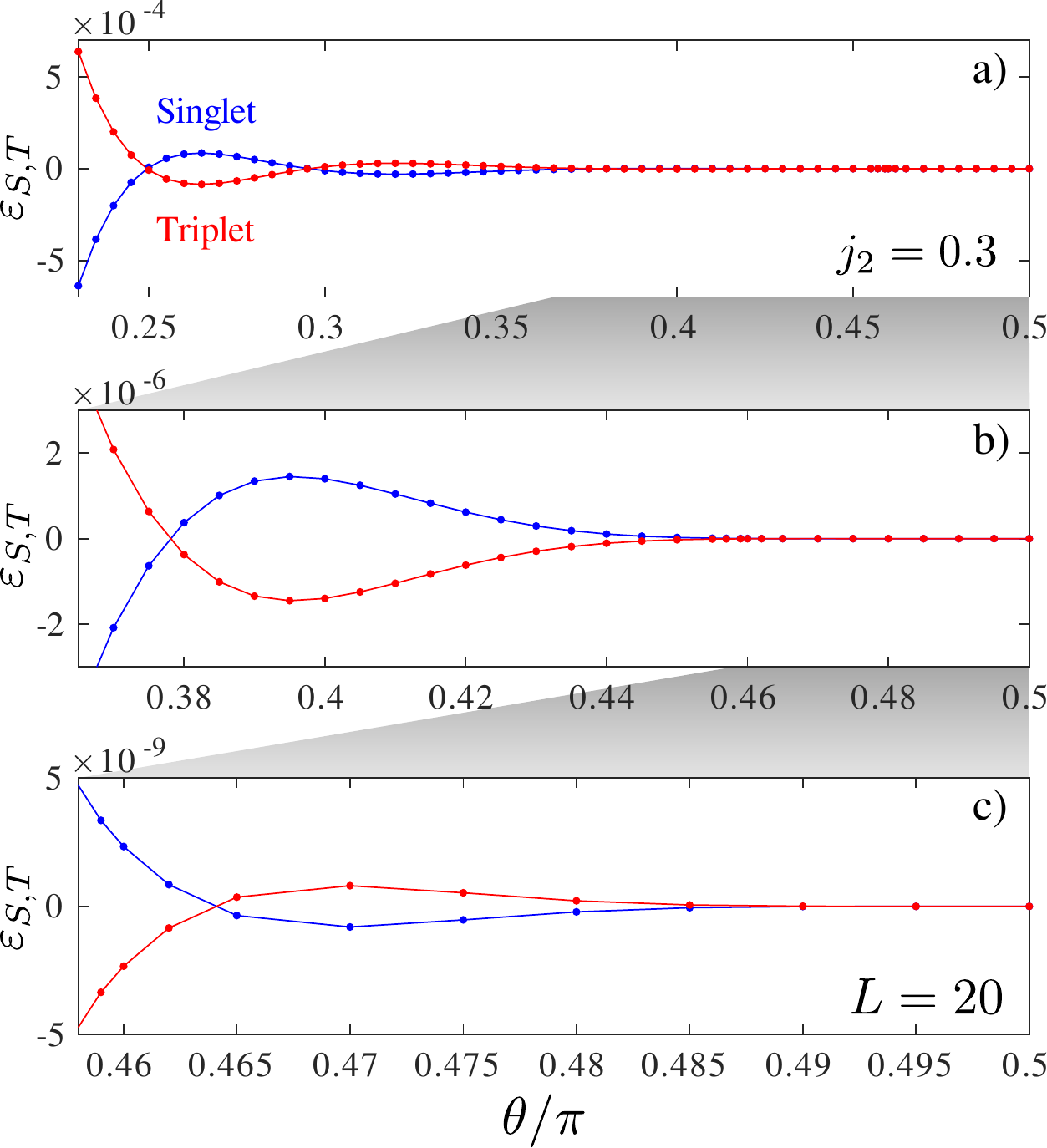}
\caption{(Color online) Multiple crossings between the singlet and triplet low-lying energy levels of the frustrated spin-1/2 ladder of Eq.(\ref{eq:ladder}) for L=20 and $j_2=0.3$ as a function of the $\theta=\arctan(J_r/J_1)$. (b), and (c) are enlarged parts of (a). }
\label{fig:energy_ladder}
\end{figure}

\begin{figure}[h!]
\includegraphics[width=0.49\textwidth]{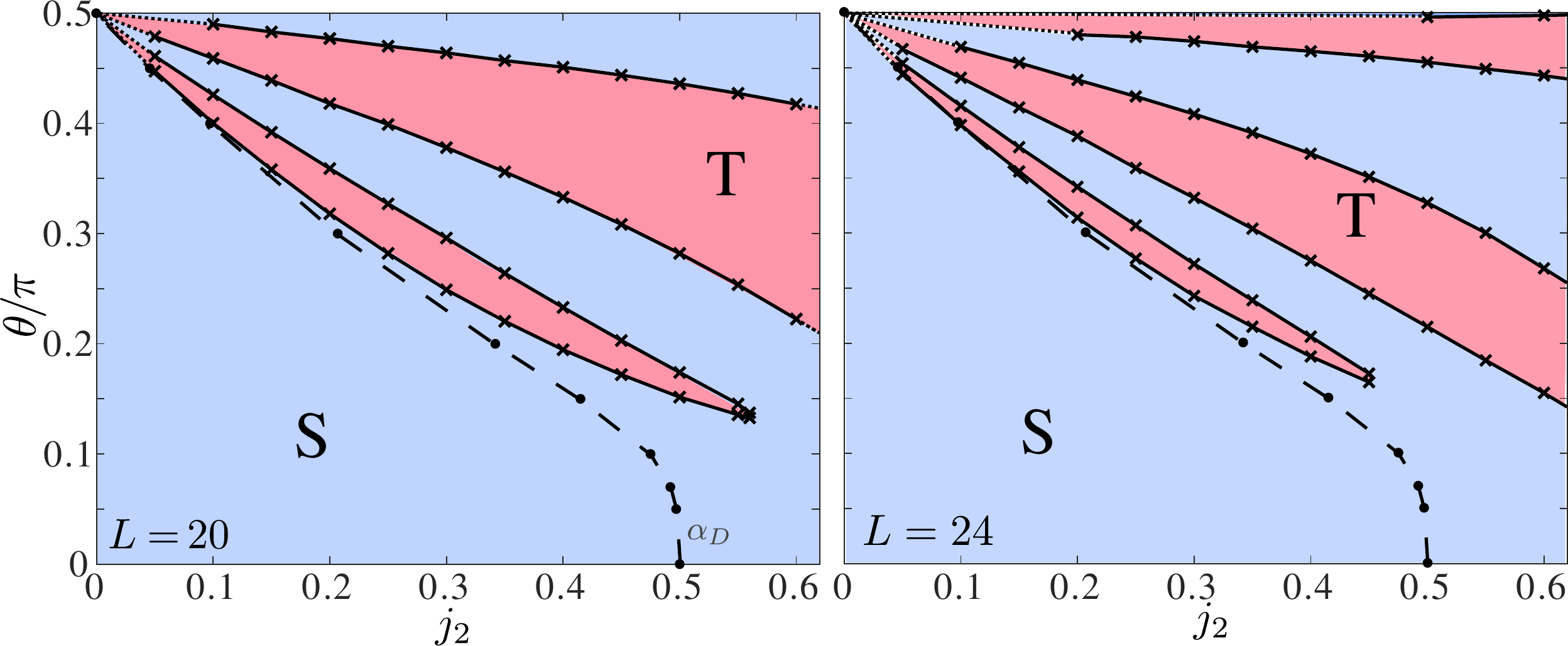}
\caption{(Color online) Singlet-triplet ground-state diagram for the spin-1/2 frustrated two-leg ladder of Eq.(\ref{eq:ladder}) with diagonal edges and $L=20$ (left) and $L=24$ (right) spins. Blue (red) areas stand for singlet (triplet) ground states with a triplet (singlet) low-lying excitation. The lower limit $\theta=0$ corresponds to a decoupled pair of spin-1/2 chains, the ground state of which is always a singlet.
For the upper limit $\theta=\pi/2$ the ground-state corresponds to decoupled dimers and spins 1/2 at the edges, so that the singlet and triplet states are always degenerate. For $L=24$ the gap for $\theta\approx \pi/2$ is below the machine precision. The dashed line stands for the disorder line.}
\label{fig:pd_ladder}
\end{figure}

Using exact diagonalizations, we find that the singlet and triplet states cross several times as a function of $\theta$ and $j_2$. The energy splitting and the level crossings as a function of $\theta$ for a fixed value of $j_2$ are illustrated in Fig.\ref{fig:energy_ladder}, while
Fig.\ref{fig:pd_ladder} summarizes our results for the singlet-triplet ground-state diagram for two different system sizes. For small next-nearest-neighbor interaction, a small change in the rung coupling can tune multiple level crossings between singlet and triplet. Experimentally this could be achieved by applying pressure along the rungs.


\section{Frustrated spin-2 chain}
\label{sec:spin2}

Until now we have only studied systems with localized spin-1/2 edge states.
Let us now generalize the concept of exact zero modes to systems with higher edge states.
Perhaps the simplest example of such a system is the spin-2 Heisenberg chain. In order to induce incommensurate correlations in the Haldane spin-2 phase, the system has to be frustrated, for instance by the next-nearest-neighbor interaction. The Hamiltonian of the model is given by:

\begin{equation}
H=\sum_{i=1}^{L-1}J_1{\bf S}_{i}\cdot{\bf S}_{i+1}+\sum_{i=1}^{L-2}J_2{\bf S}_{i}\cdot{\bf S}_{i+2},
\label{eq:j1j2}
\end{equation}
where $J_1$ and $J_2$ are both antiferromagnetic:

\begin{figure}[h!]
\includegraphics[width=0.35\textwidth]{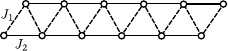}
\caption{(Color online) Sketch of the spin chain of Eq.(\ref{eq:j1j2}) with nearest- and next-nearest-neighbor interactions.}
\label{fig:zigzag_model}
\end{figure}

Without loss of generality we set $J_1=1$ throughout this section. For small $J_2$, the system is in the uniform Haldane phase with two valence-bond-singlets per nearest-neighbor bond and localized spin-1 edge states.\cite{haldane, roth_schollwoeck} These two edge spins are coupled together and form three quasi-degenerate energy levels - a singlet, a triplet and a quintuplet. As for the spin-1 chain, these states are separated by energy gaps that vanish exponentially with the chain length. 
 The correlations are incommensurate beyond the disorder point $J_2\approx0.289 $ and edge states disappear around $J_2\approx 0.46$ \cite{roth_schollwoeck}. 
 
\begin{figure}[h!]
\includegraphics[width=0.47\textwidth]{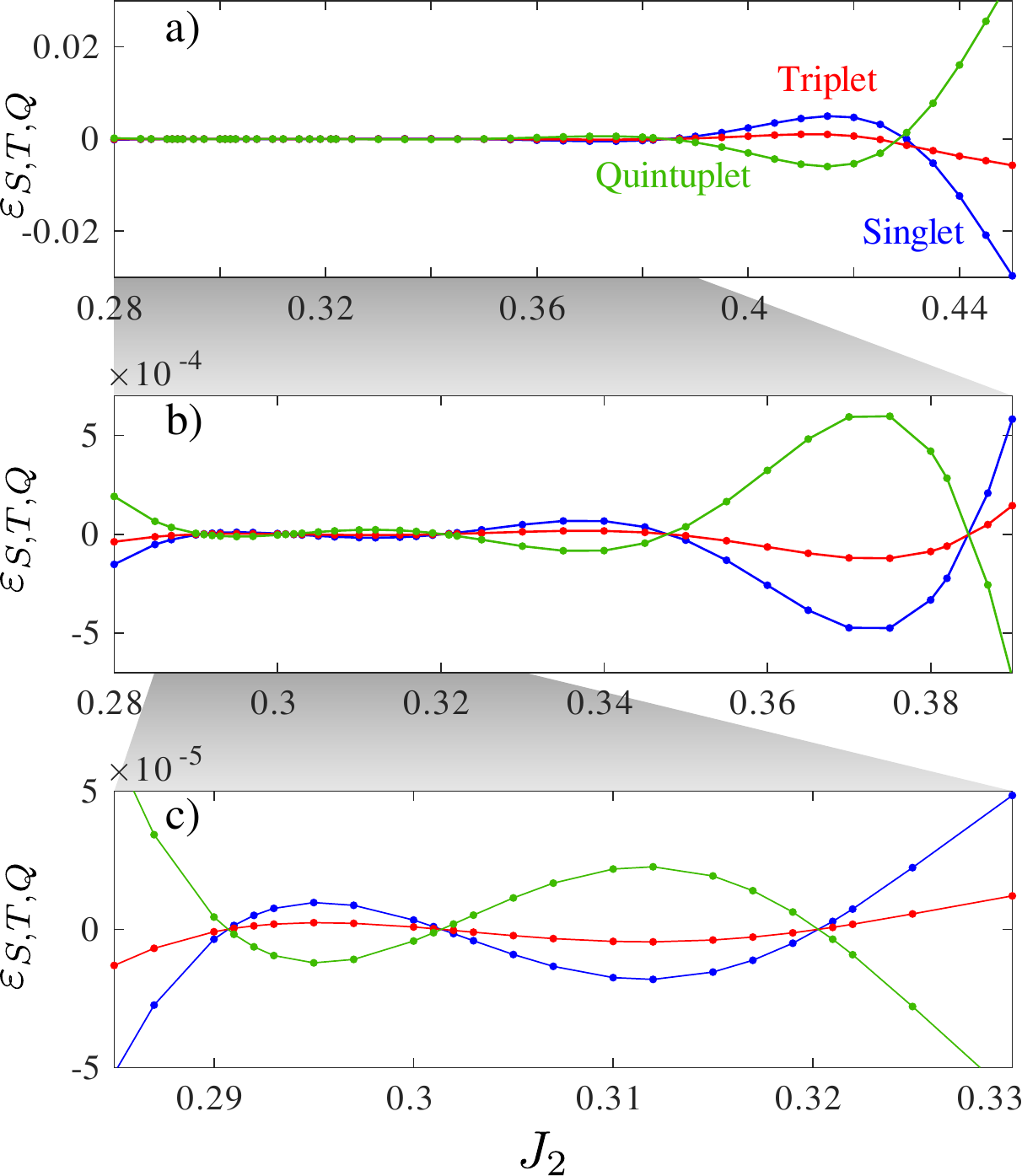}
\caption{(Color online) Multiple crossings between singlet, triplet and quintuplet low-lying energy levels for spin-2 chain with L=24 sites as a function of the next-nearest-neighbor coupling constant $J_2$. (b), and (c) are enlarged parts of (a). All three levels cross at the same point except for $J_2>0.42$, where the finite-size effects are significant.}
\label{fig:energy_s2}
\end{figure}

As above we re-scale the energies around their average:
\begin{equation}
  \varepsilon_{S,T,Q}=E_{S,T,Q}-\frac{E_S+E_T+E_Q}{3}.
\end{equation}
Fig.\ref{fig:energy_s2} shows the multiple crossings between singlet, triplet and quintuplet states in the window $0.289\leq J_2\leq 0.46$, where the Haldane phase with localized edge states is incommensurate\cite{roth_schollwoeck}.

\begin{figure}[h!]
\includegraphics[width=0.4\textwidth]{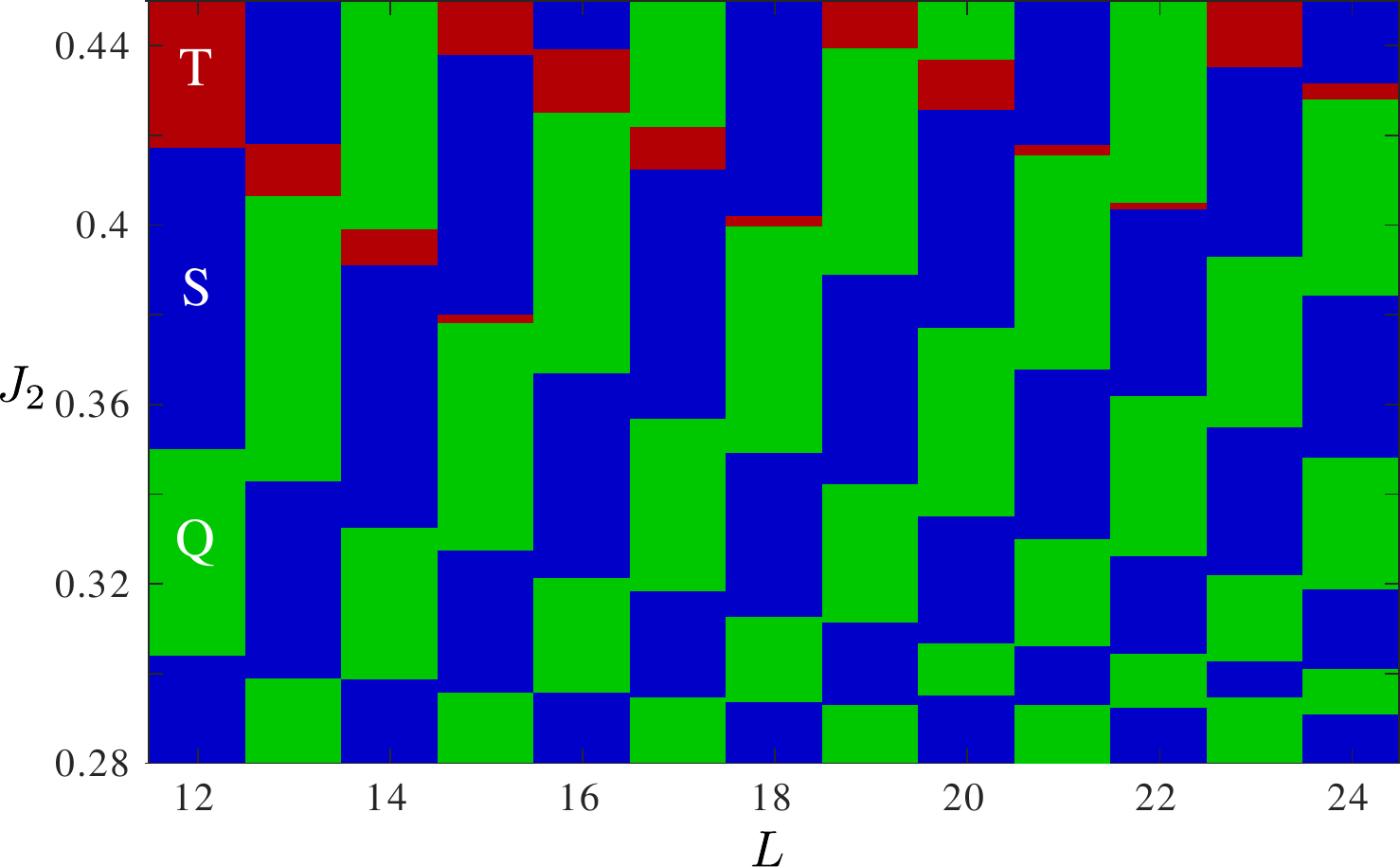}
\caption{(Color online) Phase diagram of an open spin-2 chain as a function of $J_2$ and various system sizes. Blue (green) areas stand for singlet (quintuplet) ground states with triplet and quintuplet (singlet) low-lying states. Red areas stand for regions where the ground state is a triplet. The lower limit $J_2=0.28$ lies in the commensurate Haldane phase, where the ground state simply alternates between singlet and quintuplet as a function of the size of the system.}
\label{fig:size_depend_s2}
\end{figure}

\begin{figure}[h!]
\includegraphics[width=0.47\textwidth]{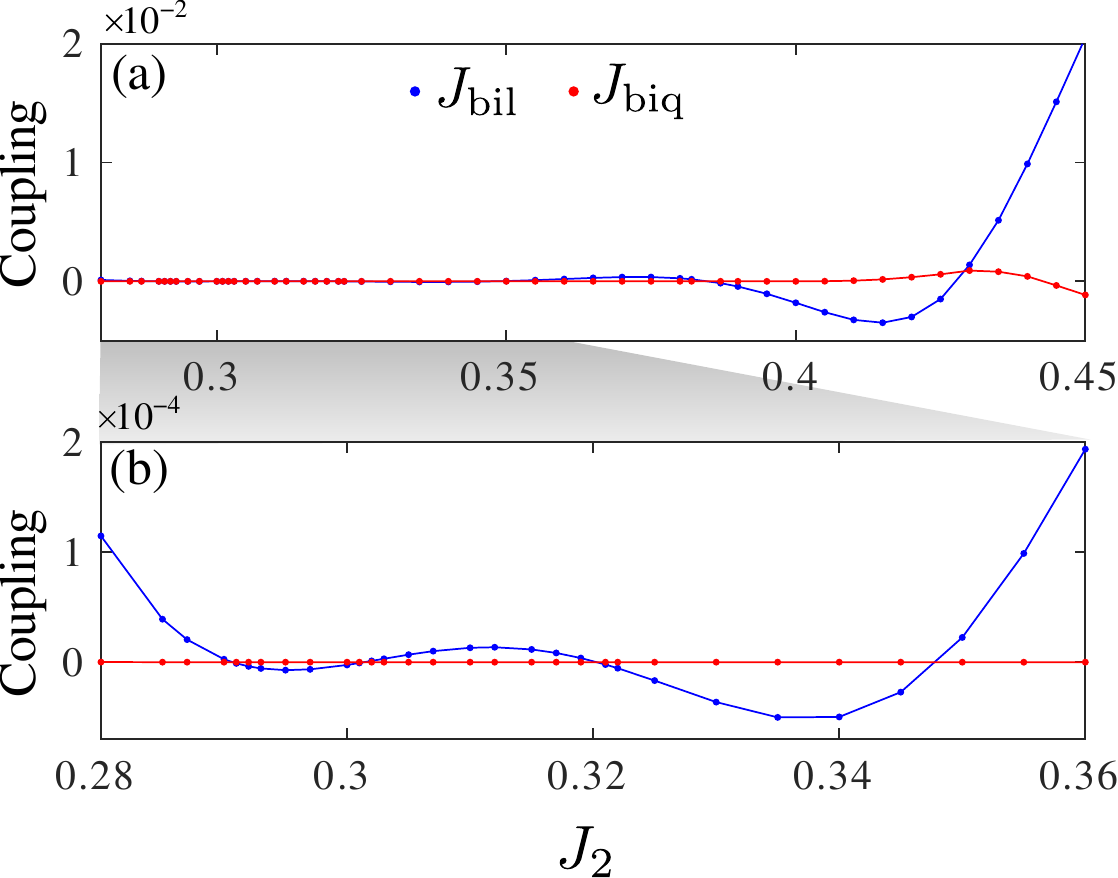}
\caption{(Color online) (a) Effective bilinear $J_\mathrm{bil}$ and biquadratic $J_\mathrm{biq}$ couplings between the edge spins. (b) is an enlarged part of (a)}
\label{fig:quadr_part}
\end{figure}

Following the general argument that the sign of the coupling is directly related to that of the spin-spin correlation between the first and last spin of the chain, we expect the edge spins to experience a simple magnetic coupling that changes sign as a function of the wave-vector of the incommensurate correlations or of the length of the system, in which case the ground state is either a singlet or a quintuplet, but never a triplet.
However, the most general, SU(2) invariant effective interaction between two spins-1 includes a biquadratic interaction on top of the bilinear one:
\begin{equation}
  H=J_\mathrm{bil}{\bf S}_1\cdot {\bf S}_2+J_\mathrm{biq}({\bf S}_1\cdot {\bf S}_2)^2.
\end{equation}
To determine the effective bilinear and biquadratic couplings between the edge spins, we note that the energy of the singlet, triplet and quintuplet states in terms of the coupling constants $J_\mathrm{bil}$ and $J_\mathrm{biq}$ are given by:
\begin{eqnarray*}
E_S&=&-2(J_\mathrm{bil}-2J_\mathrm{biq});\\
E_T&=&J_\mathrm{biq}-J_\mathrm{bil};\\
E_Q&=&J_\mathrm{bil}+J_\mathrm{biq}.
\end{eqnarray*}
Accordingly, one can extract the effective bilinear and biquadratic couplings from the low-energy spectrum according to:
\begin{eqnarray*}
J_\mathrm{bil}&=&\frac{E_Q-E_T}{2},\\
J_\mathrm{biq}&=&\frac{E_Q}{6}-\frac{E_T}{2}+\frac{E_S}{3}.
\end{eqnarray*}
The effective coupling constants (in units of $J_1$) are shown in Fig.\ref{fig:quadr_part}. The amplitude of the bilinear component is always significantly larger than the effective biquadratic coupling, as expected. In fact, the biquadratic coupling is negligible except around the very last crossing, where the correlation length is already quite large\cite{roth_schollwoeck} with respect to the system size $L=24$ so that the effective couplings take significant values. This is also illustrated in Fig.\ref{fig:size_depend_s2}: The phase diagram as a function of $J_2$ and of the chain length is dominated by an alternance of singlet and quintuplet. It is only close to the upper boundary of the phase diagram that triplet regions show up. 

\begin{figure}[h!]
\includegraphics[width=0.47\textwidth]{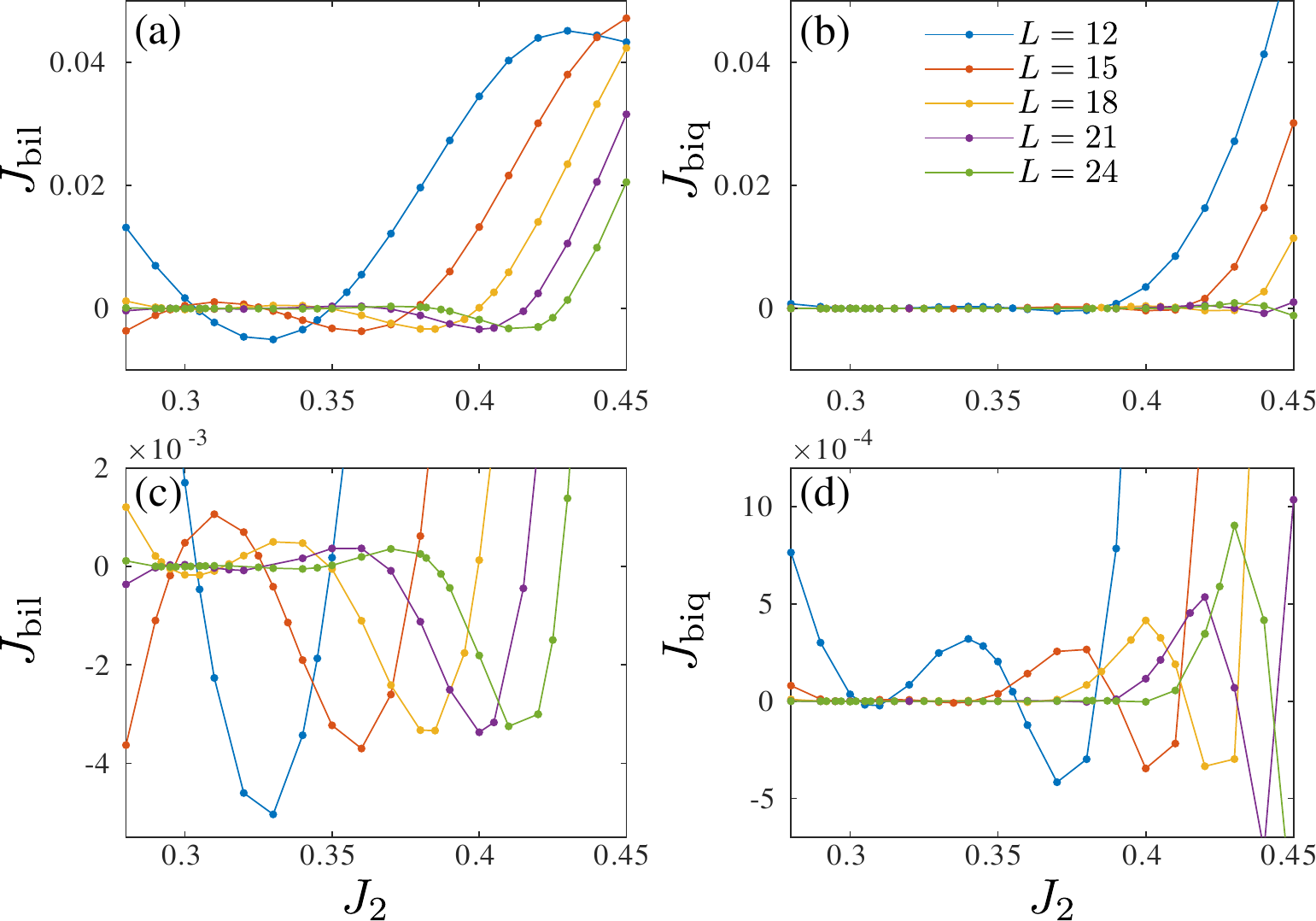}
\caption{(Color online)  Effective (a) bilinear $J_\mathrm{bil}$ and (b) biquadratic $J_\mathrm{biq}$ couplings between the edge spins for various length of the chain. (c) and (d) are enlarged parts of (a) and (b)}
\label{fig:quadr_finite_size}
\end{figure}

Another indication that the biquadratic coupling is a secondary effect comes from the analysis of the scaling of these couplings with the size of the system. As shown in Fig.\ref{fig:quadr_finite_size}, the biquadratic coupling decreases much faster with the size than the bilinear one.

As stated above, in the absence of biquadratic coupling, the  ground-state is never a triplet. It oscillates between the singlet and the quintuplet, while the first excited state is always a triplet. This could of course be changed by applying a small external magnetic field that will shift the energy of the triplet and quintuplet levels. For example, a uniform magnetic field as small as $h/J_1=-5\cdot 10^{-6}$ would allow the ground state to alternate between all three sectors.

\section{Ferromagnetic spin-1 chain}
\label{sec:fms1}

Another simple model that produces spin-1 edge states is the spin-1 chain with ferromagnetic nearest- and antiferromagnetic next-nearest-neighbor interaction. The Hamiltonian is given by Eq.\ref{eq:j1j2} with $J_1<0$, and for convenience we use the following parametrization: $J_1=\cos\theta$ and $J_2=\sin\theta$ with $\pi/2\leq \theta\leq \pi$.

\begin{figure}[h!]
\includegraphics[width=0.35\textwidth]{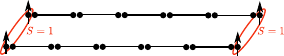}
\caption{(Color online) Valence bond singlet (VBS) picture of the ground-state in the NNN-Haldane phase with ferromagnetic NN coupling }
\label{fig:vbs_fms1}
\end{figure}

In the absence of nearest-neighbor coupling ($\theta=\pi/2$) the ground state is given by two decoupled Haldane chains. The ground-state can be visualized using valence bond singlet (VBS) at every next-nearest-neighbor bond as shown in Fig.\ref{fig:vbs_fms1}(a). The next-nearest-neighbor (NNN) Haldane phase is stabilized for $0.205\pi\leq\theta\lesssim0.87\pi$\cite{kolezhuk_prl,KolezhukPRB,chepiga_unpublished}. The full phase diagram will be reported elsewhere.
Each Haldane chain has emergent spin-1/2 edge states. In the case of antiferromagnetic nearest-neighbor interaction, the two nearest edge states, one for each of the two  Haldane chains, are coupled into a singlet, and the complete system does not have edge states. By contrast, ferromagnetic nearest-neighbor interaction leads to the formation of spin-1 edge states as shown in Fig.\ref{fig:vbs_fms1}.
 
\begin{figure}[h!]
\includegraphics[width=0.45\textwidth]{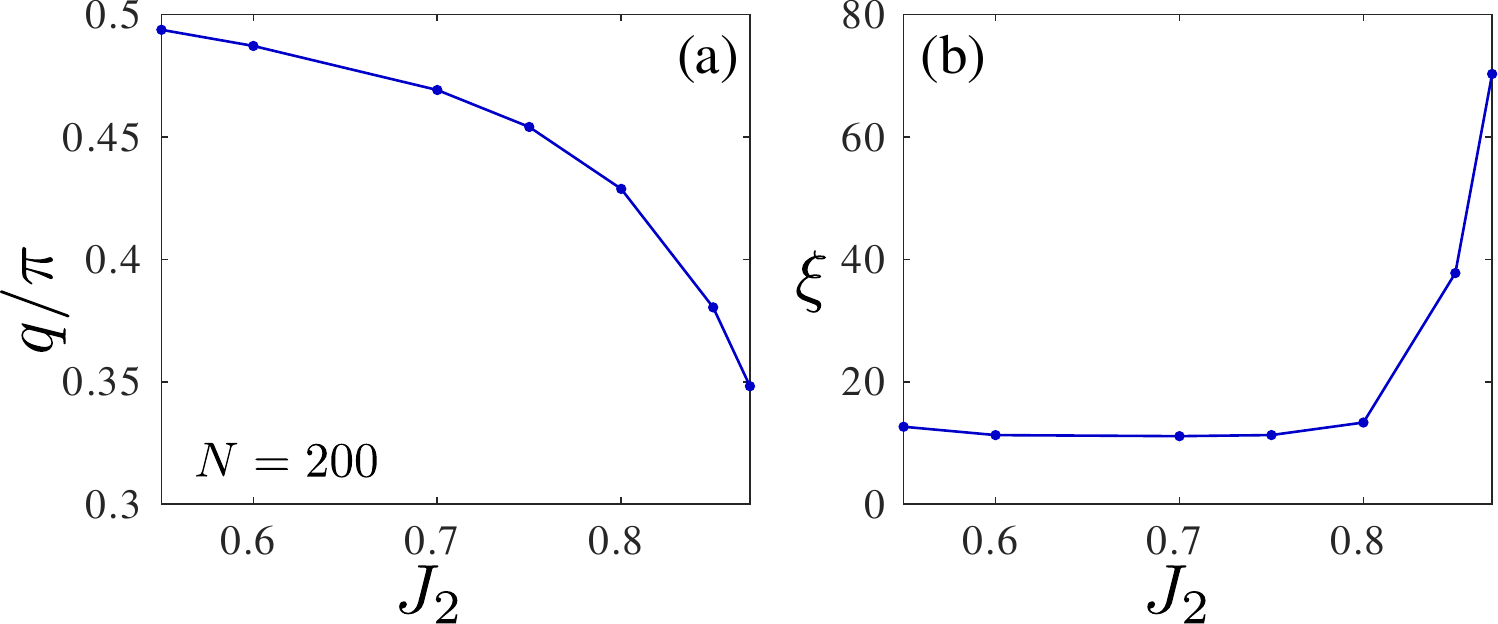}
\caption{(Color online) (a) Wave-vector and (b) correlation length as a function of $\theta$.}
\label{fig:j1j2_corel}
\end{figure}

In the presence of antiferromagnetic inter-chain coupling, it has been argued that the ground-state in the NNN-Haldane phase is given by  two intertwined VBS strings \cite{kolezhuk_connectivity}, while for ferromagnetic nearest-neighbor coupling, the correlations are incommensurate with wave-vector $0<q<\pi/2$ (see  Fig.\ref{fig:j1j2_corel}(a)), in agreement with the presence of incommensurate short range order recently reported in Ref.\onlinecite{PhysRevB.95.024424}.

\begin{figure}[h!]
\includegraphics[width=0.49\textwidth]{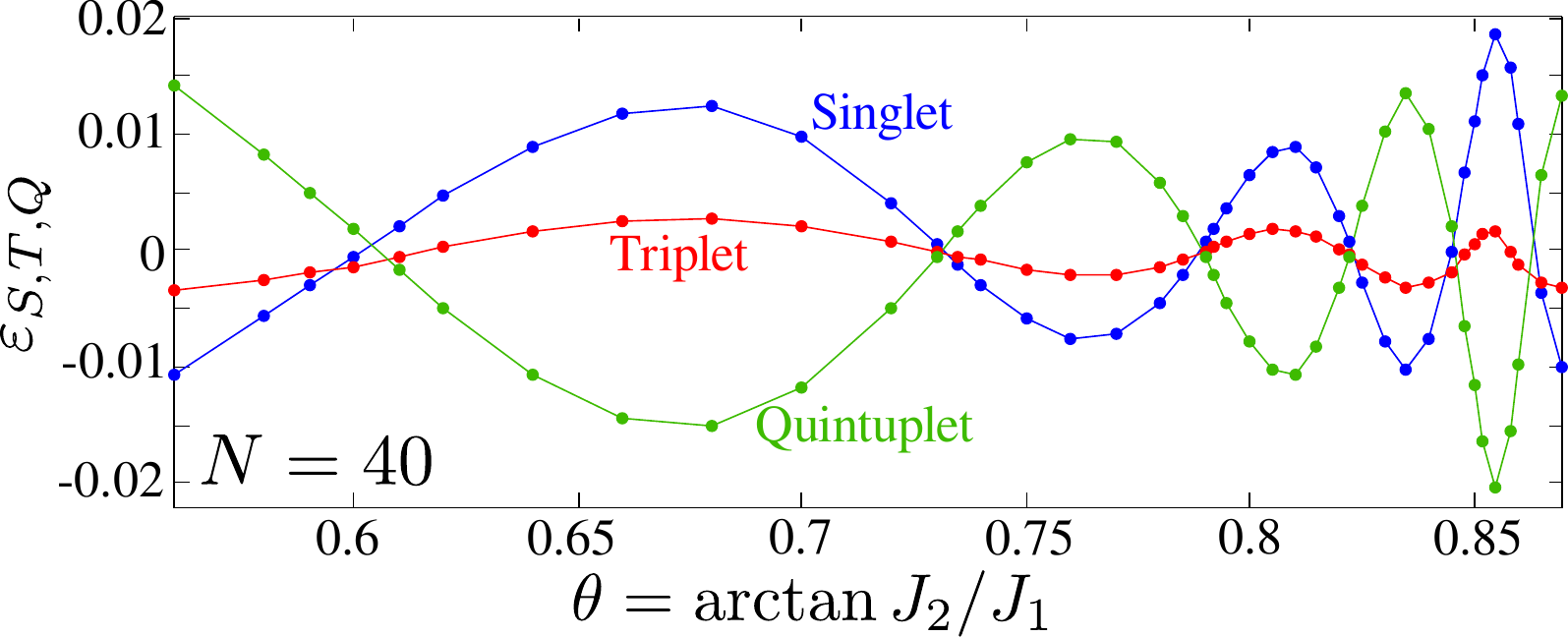}
\caption{(Color online) Multiple crossings between singlet, triplet and quintuplet low-lying energy levels for spin-1 chain with ferromagnetic nearest-neighbor and antiferromagnetic next-nearest-neighbor interactions and for $L=40$ as a function of $\theta=\arctan(J_2/J_1)$.  }
\label{fig:energy_fms1}
\end{figure}

\begin{figure}[h!]
\includegraphics[width=0.47\textwidth]{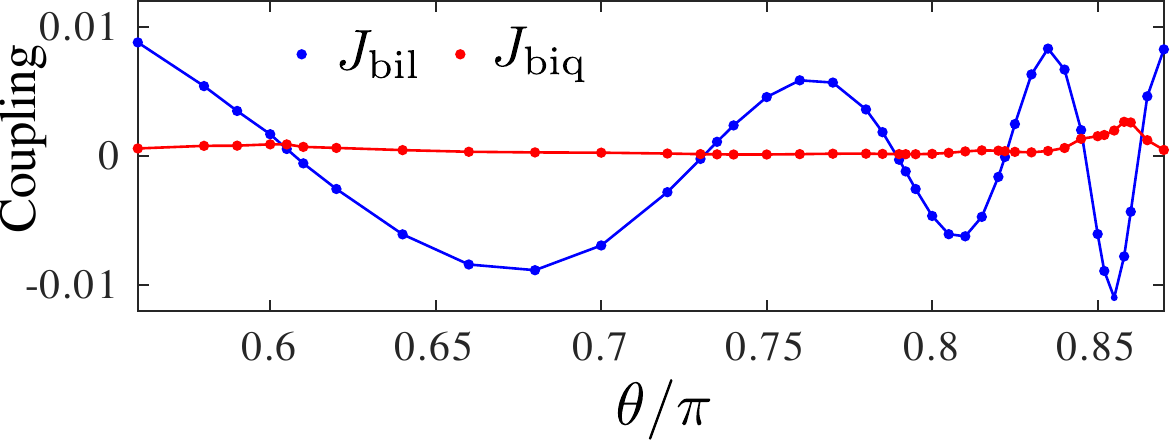}
\caption{(Color online)  Effective bilinear $J_\mathrm{bil}$ and biquadratic $J_\mathrm{biq}$ couplings between the edge spins in units of the $J_1=\cos\theta$ coupling of the original model.}
\label{fig:quadr_part_fms1}
\end{figure}

Fig.\ref{fig:energy_fms1} shows multiple crossings between the re-scaled  singlet, triplet and quintuplet low-lying in-gap states. As in the case of the spin-2 chain, all three states are almost completely degenerate at the points of exact zero modes. This implies that the biquadratic coupling between the spins-1 edge states is negligibly small in the NNN-Haldane phase, as confirmed in Fig. \ref{fig:quadr_part_fms1}.
The minor discrepancy in the last crossings is again due to the fact that the total system size is smaller that the correlation length (see Fig.\ref{fig:j1j2_corel}(b)).

Interestingly enough, due to the very large correlation length in this model, the energy difference between quasi-degenerate singlet, triplet and quintuplet states remains significant for relatively large system size ($N=40$). We hope that this will inspire further investigation towards the experimental realization of exact zero modes in the topologically non-trivial phases of spin-$S$ systems.

\section{Conclusions}
\label{sec:conclusion}

In the present paper we have shown that the appearance of points with exactly degenerate low-lying in-gap states is a generic feature of systems with {\it i)} localized edge states and {\it ii)} incommensurate correlations. The mechanism is general with respect to the value of the edge spins, of the bulk spin, and of the geometry. Besides, when the edge spins are not spins-1/2 but spins-1, we have shown that the coupling is almost purely magnetic with very small biquadratic component. These observations show that frustration leads to a simple mechanism to produce localized spins with a tunable, purely magnetic effective coupling.

\section{Acknowledgments}

We thank Ian Affleck for useful discussion on the properties of the AKLT state.
This work has been supported by the Swiss National Science Foundation.
The calculations have been performed using the facilities of the Scientific IT and Application Support Center of EPFL.

\bibliographystyle{apsrev4-1}
\bibliography{bibliography}

\end{document}